\documentstyle[namedreferences]{kapproc}

\newcommand{\al}{\alpha}

\newcommand{\bb}{\begin{equation}}

\newcommand{\be}{\beta}
\newcommand{\bega}{\begin{eqnarray}}
\newcommand{\begae}{\begin{eqnarray*}}

\newcommand{\bwe}{\mbox{$\bigwedge$}}
\newcommand{\C}{\;\mbox{{\sf I}}\!\!\!C}

\newcommand{\dis}{\displaystyle}

\newcommand{\divi}{{\rm div}}

\newcommand{\ee}{\end{equation}}
\newcommand{\ega}{\end{eqnarray}}
\newcommand{\egae}{\end{eqnarray*}}

\newcommand{\ga}{\gamma}

\newcommand{\h}{\hspace*{7mm}}

\newcommand{\ia}{{\bf i}}
\newcommand{\infi}{\infty}

\newcommand{\lan}{\langle}
\newcommand{\ld}{\ldots}
\newcommand{\leri}{\leftrightarrow}

\newcommand{\nab}{\nabla}
\newcommand{\nb}{\nonumber}

\newcommand{\Om}{\Omega}
\newcommand{\om}{\omega}

\newcommand{\ot}{\otimes}
\newcommand{\ov}{\overline}

\newcommand{\pa}{\partial}

\newcommand{\R}{I\!\!R}

\newcommand{\ro}{\mbox{\boldmath $\rho$}}
\newcommand{\ran}{\rangle}

\newcommand{\rot}{{\rm rot}}

\newcommand{\sub}{\subset}
\newcommand{\ven}{{\bf v}}

\newcommand{\we}{\wedge}

\newcommand{\thetao}{\mbox{\boldmath $\theta$}}
\newcommand{\wid}{\widetilde}
\newcommand{\zo}{\mbox{\boldmath $z$}}

\newcommand{\clif}{{\cal C}\!\ell}
\newcommand{\dirac}{{\mbox{\boldmath$\partial$}}}
\newcommand{\proje}{{\mbox{\boldmath$\Lambda$}}}
\newcommand{\prospin}{{\mbox{\boldmath$\Sigma$}}}

\makeatletter
\@addtoreset{equation}{section}
\makeatother

\renewcommand{\theequation}{\arabic{section}.\arabic{equation}}

\runningtitle{SUBLUMINAL AND SUPERLUMINAL ELECTROMAGNETIC WAVES \ldots}

\begin{opening}

\title{Subluminal and Superluminal Electromagnetic Waves and the Lepton
Mass Spectrum}

\author{Waldyr Alves Rodrigues, Jr.\thanks{walrod@ime.unicamp.br}}

\institute{Instituto de Matem\'atica, Estat\'{\i}stica e Ci\^encia da
Computa\c{c}\~ao \\
IMECC - UNICAMP, CP 6065 \\
13081-970, Campinas, SP, Brasil}

\author{Jayme Vaz, Jr.\thanks{vaz@suhep.phy.syr.edu}}

\institute{Department of Physics \\
201 Physics Building - Syracuse University \\
Syracuse, NY, 13244-1130  USA }
\end{opening}

\begin{document}

\begin{abstract}
Maxwell equation $\dirac    F = 0$ for $F \in \sec \bwe^2 M \subset
\sec \clif (M)$, where $\clif (M)$ is the Clifford bundle of
differential forms, have subluminal and superluminal solutions
characterized by $F^2 \neq 0$. We can write $F = \psi \gamma_{21}
\tilde \psi$ where $\psi \in \sec \clif^+(M)$. We can show that $\psi$
satisfies a non linear Dirac-Hestenes Equation (NLDHE). Under
reasonable assumptions we can reduce the NLDHE to the linear
Dirac-Hestenes Equation (DHE). This happens for constant values of the
Takabayasi angle ($0$ or $\pi$).
The massless Dirac equation $\dirac  \psi =0$, $\psi \in  \sec \clif^+
(M)$, is equivalent to a generalized Maxwell equation $\dirac F = J_{e}
- \gamma_5 J_{m} = {\cal J}$.  For $\psi = \psi^\uparrow$ a positive
parity eigenstate, $j_e = 0$. Calling $\psi_e$ the solution
corresponding to the electron, coming from $\dirac F_e =0$, we show
that the NLDHE for $\psi$  such that $\psi  \gamma_{21} \tilde{\psi} =
F_e + F^{\uparrow}$  gives a linear DHE for Takabayasi angles $\pi/2$
and $3\pi/2$ with the muon mass. The Tau mass can also be obtained with
additional hypothesis.
\end{abstract}

\section{Introduction}

In section 1 we briefly recall how to write Maxwell and Dirac equations
in the Clifford and Spin-Clifford bundle formalisms. In section 2 we
present some mathematical preliminaries. Then in section 3 we prove
that the free Maxwell equations $\dirac    F = 0$,  $F \in \sec \bwe^2
M \subset \sec \clif (M)$, have subluminal and superluminal solutions
characterized by $F^2 \neq 0$. In particular, we show that there are
some interesting solutions of  $\dirac    F = 0$ which are equivalent
to a non-homogeneous  Maxwell equation $\dirac    F' = j$. In section 4
we prove that the solutions of  $\dirac    F = 0$ for which $F^2 \neq
0$ are equivalent to solutions of a non-linear Dirac-Hestenes equation
(NLDHE) for $\psi \in \sec \clif^+(M)$ such that $F = \psi \gamma_{21}
\tilde\psi$. Under reasonable assumptions, namely that $\psi$ has  only
six degrees of freedom, the NLDHE gives a linear Dirac-Hestenes
equation (DHE) for a constant mass. We may identify this solution
$\psi_e$ with the DHE for the electron (or positron) depending
on the value of the Takabayasi angle. In section 5 we study some
properties of the linear and nonlinear Dirac-Hestenes equations.
In this formalism the Weyl equation is written $\dirac \psi_W
=0$, with $\psi_W \in \sec \clif^{+} (M)$ and $\psi_W
\gamma_{21} \psi_W =0$. The expression $\dirac \psi_W =0$ is
equivalent to a generalized Maxwell equation $\dirac F = J_e -
\gamma_5 J_m = {\cal J}$. If $\psi^\uparrow$ is a Dirac-Hestenes
spinor which is a positive parity eigenstate (see eq.(2.16)) and
$\psi^\uparrow$ satisfies $\dirac \psi^\uparrow =0$, then the
equivalent Maxwell equation reads $\dirac F^\uparrow = -
\gamma_5 J_m$ (see eq.(5.42)). In section 6 we study the NLDHE
associated to $\dirac (F_e + F^\uparrow )= \dirac F = {\cal J}$,
which yields, for certain values of the Takabayasi angle
associated to $\psi$ such that $F= \psi \gamma_{21} \psi$, a DHE
with the correct value of the muon mass. Under additional
hipothesis this theory gives also the values of the Tau lepton
mass.

\section{Mathematical Preliminaries}

Here we briefly recall how to write Maxwell and Dirac equations in the
Clifford and Spin-Clifford bundles over Minkowski spacetime. Details
concerning these theories can be found in
\cite{warqui,rodsou94,wetal95a,wetal95b}.


Let ${\cal M}=(M,g,D)$ be Minkowski spacetime. $(M,g)$ is a four
dimensional
time oriented and space oriented Lorentzian manifold, with $M\simeq
{\sl I \!\! R}^4$ and $g \in {\rm sec}(T^*M \times T^*M)$ being a
Lorentzian metric
of signature (1,3). $T^*M$ [$TM$] is the cotangent [tangent] bundle.
$T^*M = \cup_{x\in
M} T^*_xM$ and $TM = \cup_{x\in M}T_xM$, and $T_xM \simeq T^*_xM \simeq
{\sl I \!\! R}^{1,3}$, where ${\sl I \!\! R}^{1,3}$ is the Minkowski
vector space \cite{REF-3,REF-3a}.
$D$ is the Levi-Civita connetion of $g$, i.e., $Dg=0$,
$\mbox{\boldmath $T$}(D) =0$. Also
$\mbox{\boldmath $R$}(D)=0$, $\mbox{\boldmath $T$}$ and
$\mbox{\boldmath $R$}$ being
respectively the torsion and curvature
tensors. Now, the Clifford bundle of differential forms $\clif(M)$ is
the bundle of algebras $\clif(M) = \cup_{x\in M} \clif (T^*_xM)$, where
$\forall x\in M, \clif(T^*_xM) = \clif_{1,3}$, the so
called spacetime
algebra \cite{REF-4,REF-4a,REF-6}. Locally as a linear space over the
real field
${\sl I \!\! R}$, $\clif(T^*_x(M))$ is isomorphic to the Cartan
algebra $\bigwedge (T^*_x(M)$ of the
cotangent space  and
$\bigwedge(T^*_x M) = \sum^4_{k=0} \bigwedge {}^k(T^*_x M)$,
where $\bigwedge^k(T^*_x M)$ is the $4 \choose k$-dimensional space of
$k$-forms. The
Cartan bundle $\bigwedge(M) = \cup_{x\in M} \bigwedge(T^*_x M)$ can
then be
thought as
``imbedded" in $\clif(M)$. In this way sections of $\clif(M)$ can be
represented as a sum of inhomogeneous differential forms. Let $\{
e_\mu = \frac{\partial}{\partial x^\mu}\} \in {\rm sec} TM, (\mu =
0,1,2,3)$ be an orthonormal basis
$g(e_\mu, e_\nu) = \eta_{\mu\nu} = {\rm diag}(1,-1,-1,-1)$ and let $\{
\gamma^\nu = d x^\nu \} \in
{\rm sec} \bigwedge^1(M) \subset {\rm sec} \clif(M)$ be the
dual basis. Then, the
fundamental Clifford product (in what follows to be denoted by
juxtaposition of symbols) is generated by
$\gamma^\mu\gamma^\nu+\gamma^\nu\gamma^\mu = 2\eta^{\mu\nu}$
and if ${\cal C} \in {\rm sec}\clif(M)$ we have

\begin{equation} \label{2e1}
{\cal C} = s+ v_\mu \gamma^\mu + \frac{1}{2!} b_{\mu\nu}
\gamma^\mu\gamma^\nu +
\frac{1}{3!} a_{\mu\nu\rho} \gamma^\mu\gamma^\nu\gamma^\rho + p
\gamma^5 \; ,
\end{equation}
where $\gamma^5 = \gamma^0\gamma^1\gamma^2\gamma^3
= dx^0 dx^1 dx^2 dx^3$  is the volume element and
$s,v_\mu, b_{\mu v}, a_{\mu\nu\rho}, p\in {\rm sec} \bigwedge^0(M)
\subset {\rm sec}
\clif(M)$. For $A_r \in {\rm sec} \bigwedge^r(M)\subset {\rm sec}
{\cal C}(M), B_s \in {\rm sec}
\bigwedge^s(M)$ we define \cite{REF-4,lo94,REF-4a}
$A_r\cdot B_s = \langle A_rB_s \rangle_{|r-s|}$ and
$A_r\wedge B_s = \langle A_rB_s \rangle_{r+s}$,
where $\langle \hspace{1ex} \rangle_k$ is the component in
$\bigwedge^k(M)$ of the Clifford field.

Besides the vector bundle $\clif(M)$ we need
also to introduce another vector
bundle $\clif_{{\rm Spin}_+(1,3)}(M)$
$[{\rm Spin}_+(1,3) \simeq
{\rm SL}(2,{\sl I \!\!\!\! C})]$
called the Spin-Clifford bundle \cite{rodfig90}. We can show that
$\clif_{{\rm Spin}_+(1,3)}(M)\simeq \clif(M)/{\cal R}$,
i.e., it is a quotient
bundle. This means that sections of $\clif_{{\rm Spin}_+(1,3)}(M)$ are
some special equivalence classes of sections of the Clifford bundle,
i.e, they are equivalence sections of non-homogeneous differential
forms (see eqs.(\ref{eq.1},\ref{eq.2}) below).

Now, as is well known, an electromagnetic field is represented by $F
\in
{\rm sec} \bigwedge^2(M) \subset {\rm sec} \clif(M)$. How to
represent the Dirac spinor
fields in this formalism~? We can show that even sections of
$\clif_{\mbox{Spin}_+(1,3)}(M)$, called Dirac-Hestenes spinor fields,
do the
job. If we fix two orthonormal basis, $\Sigma = \{\gamma^\mu\}$ as
before,
and
$\dot{\Sigma} = \{\dot{\gamma}^\mu = R\gamma^\mu \widetilde{R} =
\Lambda^\mu_\nu \gamma^\nu \}$ with $\Lambda^\mu_\nu \in
{\rm SO}_{+}(1,3)$ and
$R\in {\rm Spin}_+(1,3) \; \forall x \in M$, $R\widetilde{R} =
\widetilde{R} R =1$,
and where
$\, \widetilde{} \, $ is the reversion operator in $\clif_{1,3}$
\cite{REF-4,lo94,REF-4a}, then the
representations of an even section $\mbox{\boldmath $\psi$} \in {\rm
sec}
\clif_{{\rm Spin}_+(1,3)}(M)$ are the sections $\psi_\Sigma$ and
$\psi_{\dot{\Sigma}}$ of $\clif(M)$ related by
\begin{equation}
\label{eq.1}
\psi_{\dot\Sigma} = \psi_\Sigma R
\end{equation}
and
\begin{equation}
\label{eq.2}
\psi_\Sigma = s + \frac{1}{2!} b_{\mu\nu} \gamma^\mu \gamma^\nu + p
\gamma^5
\end{equation}
Note that $\psi_{\Sigma}$ has the correct number of degrees of
freedom in order to represent a Dirac spinor field, which is not the
case with the so called Dirac-K\"ahler spinor field.

Let $\star$ be the Hodge star operator $\star :\bigwedge^k(M)
\rightarrow
\bigwedge^{4-k}(M)$.
Then we can show that if $A_p \in {\rm sec} \bigwedge^p(M)
\subset {\rm sec}
\clif(M)$ we have
$\star A = \widetilde{A} \gamma^5$.
Let $d$ and $\delta$ be respectively the differential and Hodge
codifferential operators acting on sections of $\bigwedge(M)$. If
$\omega_p \in {\rm sec}
\bigwedge^p(M)\subset {\rm sec} \clif(M)$, then $\delta \omega_p
= (-)^p \star^{-1} d \star \omega_p$, with $\star^{-1}\star = {\rm
identity}$.

The Dirac operator acting on sections of $\clif(M)$ is the invariant
first order differential operator
\begin{equation}
\dirac = \gamma^\mu D_{e_{\mu}} ,
\end{equation}
and we can show the very important result \cite{REF-5}:
\begin{equation}
\dirac = \dirac \wedge \,  + \, \dirac \cdot = d-\delta .
\end{equation}

With these preliminaries we can write Maxwell and Dirac equations  as
follows \cite{REF-6,rodoli90}:
\begin{equation}
\dirac F = 0 ,
\end{equation}
\begin{equation} \label{2e7}
\dirac \psi_{\Sigma} \gamma^1\gamma^2 + m \psi_\Sigma \gamma^0 =0 .
\end{equation}
If $m=0$ we have the massless Dirac equation
\begin{equation}
\label{eq.10}
\dirac \psi_\Sigma = 0 ,
\end{equation}
which is Weyl's equation (see eq.(\ref{2e12n}) below) when
$\psi_\Sigma$ is reduced to a Weyl spinor field \cite{REF-4,lo94}.
Note that in this formalism Maxwell equations condensed in a single
equation!
Also, the specification of $\psi_\Sigma$ depends on the frame
$\Sigma$. When no confusion arises we represent $\psi_\Sigma$ simply by
$\psi$.

When $\psi_\Sigma \tilde{\psi_\Sigma} \neq 0$, where $\sim$ is the
reversion operator, then $\psi_\Sigma$ has the following cannonical
decomposition:

\begin{equation}
\psi_\Sigma = \sqrt{\rho} e^{\beta\gamma_5/2} R \, ,
\end{equation}
where $\rho$, $\beta \in \sec \bwe^0 (M) \subset \sec \clif(M)$ and $R
\in \mbox{Spin}_+(1,3) \subset \clif^{+}_{1,3}$, $\forall x \in M$.
$\beta$ is called the Takabayasi angle.

If one wants to work in terms of the usual spinor field formalism,
we can translate our results by choosing, for example,
the standard matrix representation of $\{\gamma^\mu\}$, and
for $\psi_{\Sigma}$ given by eq.(\ref{eq.2}) we have the
following (standard) matrix representation \cite{wetal95b}:

\begin{equation}
\Psi = \left( \begin{array}{cc}
	      \phi_1 & -\phi_2^* \\
	      \phi_2 & \phi_1^*
	      \end{array} \right) ,
\end{equation}
where
\begin{equation}
\phi_1 = \left( \begin{array}{cc}
		s - ib_{12} & b_{13}-ib_{23} \\
		-b_{13}-ib_{23} & s+ib_{12}
		\end{array} \right) , \quad
\phi_2 = \left( \begin{array}{cc}
		-b_{03}+i p & -b_{01}+ib_{02} \\
		-b_{01} - ib_{02} & b_{03}+ i p
		\end{array} \right) .
\end{equation}
with $s, \, b_{12}, \ldots$ real functions.
Right multiplication by
$$
\left( \begin{array}{c}
	  1 \\ 0 \\ 0 \\ 0
       \end{array} \right)
$$
gives the usual Dirac spinor field.


We need also the concept of Weyl spinors. By definition, $\psi \in \sec
\clif^+ (M)$ is a Weyl spinor if (\citeauthor{REF-4}
\citeyear{REF-4},\citeyear{lo94})

\begin{equation} \label{2e12n}
\gamma_5 \psi = \pm \psi \gamma_{21} \; .
\end{equation}
The positive (negative) ``eigenstate'' of $\gamma_5$ will be denoted
$\psi_+$ ($\psi_-$). For a general $\psi \in \sec \clif^+ (M)$ we can
write

\begin{equation}
\psi_\pm = \frac{1}{2} [ \psi \mp \gamma_5 \psi \gamma_{21} ]
\end{equation}
and then

\begin{equation}
\psi = \psi_+ + \psi_- \; .
\end{equation}

The parity operator $P$ in this formalism is represented in such a way
(\citeauthor{REF-4} \citeyear{REF-4},\citeyear{lo94}) that for $\psi
\in \sec \clif^+ (M)$,

\begin{equation}
P \psi = - \gamma_0 \psi \gamma_0 \; .
\end{equation}

The following Dirac-Hestenes spinors are eigenstates of the parity
operator with eigenvalues $\pm 1$:

\begin{equation} \label{2e17}
\begin{array}{c}
P \psi^\uparrow = + \psi^\uparrow \; ,  \quad \psi^\uparrow = \gamma_0
\psi_- \gamma_0 - \psi_- \; ; \\
P \psi^\downarrow = - \psi^\downarrow \; ,  \quad \psi^\downarrow =
\gamma_0 \psi_+ \gamma_0 + \psi_+ \; .
\end{array}
\end{equation}


We recall that the even subbundle  $\clif^+(M)$ of $\clif(M)$ is such
that its typical fiber is the Pauli algebra $\clif_{3,0} \equiv
\clif_{1,3}^{+}$. The isomorphism $\clif_{3,0} \equiv \clif_{1,3}^{+}$
is exhibited by putting $\sigma_i = \gamma_i \gamma_0$, whence
$\sigma_i \sigma_j + \sigma_j \sigma_i = 2 \delta_{ij}$. Then if
$\displaystyle F = \frac{1}{2} F^{\mu\nu} \gamma_\mu \gamma_\nu \in
\sec \bwe^2M \subset \sec \clif(M)$ with

\begin{equation} \label{2e12}
F^{\mu\nu} =
\left(
\begin{array}{cccc}
0 & -E^1 & -E^2 & - E^3 \\
E^1 &  0 & -B_3  &  B_2 \\
E^2 &  B_3 & 0 & -B_1 \\
E^3 &  -B_2 & B_1 & 0
\end{array}
\right)  \; ,
\end{equation}
we can write $F = \vec{E} + \mbox{\bf i} \vec{B}$ with $\mbox{\bf i} =
\sigma_1 \sigma_2 \sigma_3 = \gamma_5$, $\vec{E} = E^i \sigma_i$,
$\vec{B} = B^i \sigma_i$.

Before we present the subluminal and superluminal
solutions $F_<$ and $F_>$  of Maxwell equations we shall define
precisely an inertial reference
frame (irf) \cite{REF-3,REF-3a}. An irf $I\in {\rm sec} TM$ is a
timelike
vector
field pointing into the future such that $g(I,I) =1$ and $DI=0$. Each
integral line of $I$ is called an observer. The coordinate functions
$\langle x^\mu
\rangle$ of a chart of
the maximal atlas of $M$ are called naturally adapted to $I$ if
$I=\partial/\partial x^0 $. Putting $I=e_0$, we can find $e_i =
\partial/\partial x^i$ such that
$g(e_\mu,e_\nu) = \eta_{\mu\nu}$ and the coordinate functions $\langle
x^\mu\rangle$ are the usual Einstein-Lorentz ones and have a
precise operational meaning \cite{REF-7}. $x^0$ is measured by
``ideal clocks" at rest synchronized ``\`a la Einstein'' and $x^i$
are measured with ``ideal rulers".

\section{Subluminal and Superluminal Undistorted Progressive Waves
(UPWs) Solutions of Maxwell
Equations (ME)}


We start
by reanalyzing in section 3.1 the plane wave solutions (PWS) of ME
with the Clifford bundle formalism. We clarify some misconceptions and
explain the
fundamental role of the duality operator $\ga_5$ and the meaning of
$i=\sqrt{-1}$ in standard formulations of electromagnetic theory.

Next, in section 3.2 we discuss subluminal UPWs solutions of ME and an
unexpected relation between these solutions and the possibility of the
existence of purely electromagnetic particles (PEPs) envisaged by
\cite{einstein19,poinc06,ehren07} and recently
discussed by Waite,  Barut and Zeni \cite{wai95,waietal96}.

In section 3.3 we discuss in detail the theory of superluminal
electromagnetic $X$-waves (SEXWs). In \cite{upw} it is discussed how to
produce these waves
with appropriate physical devices.

\subsection{Plane Wave Solutions of Maxwell Equations}

We recall that ME in vacuum can be written as

\bb
\dirac F = 0,
\ee
where $F \in \sec \bwe^2(M) \subset \sec \clif(M)$. The well known PWS
of
eq.(3.1) are obtained as follows. We write in a given Lorentzian chart
$\lan x^\mu\ran$ of the maximal atlas of $M$  a PWS
moving in the $z$-direction
\bb
F = f e^{\ga_5 kx}
\ee
\bb
k = k^\mu \ga_\mu, k^1=k^2=0, x = x^\mu\ga_\mu ,
\ee
where $k, x \in \sec \bwe^1(M) \sub \sec \clif(M)$ and where $f$
is a constant 2-form. From eqs(3.2) and (3.3) we obtain
\bb
kF=0
\ee
Multiplying eq(3.4) by $k$ we get
\bb
k^2F=0
\ee
and since $k \in \sec \bwe^1(M) \sub \sec \clif(M)$ then
\bb
k^2 = 0 \ \leri \ k_0 = \pm |\vec k|=k^3 ,
\ee
i.e., the propagation vector is light-like. Also
\bb
F^2 = F.F + F \we F = 0
\ee
as can be easily seen by multiplying both members of eq(3.4) by $F$ and
taking into account that $k \neq 0$. Eq(3.7) says that the field
invariants are null.

It is interesting to understand the fundamental role of the volume
element $\ga_5$ (duality operator) in electromagnetic theory. In
particular since $e^{\ga_5 kx} = \cos kx + \ga_5 \sin kx$, we see
that
\bb
F = f \cos k x + \ga_5 f \sin kx,
\ee
i.e., in a PWS the electric and magnetic fields are oscillating out of
phase by 90$^{\circ}$. Writing $F = \vec E + \ia \vec B$, (see
eq.(\ref{2e12})) with $\ia \equiv \ga_5$ and choosing $f = \vec e$,
eq.(3.8)
becomes
\bb
(\vec E + \ia\vec B) = \vec e \cos k x + \ia \vec e \sin k x .
\ee
This equation is important because it shows that we must take care
with the $i=\sqrt{-1}$ that appears in usual formulations of Maxwell
Theory using complex electric and magnetic fields. The $i=\sqrt{-1}$
in many cases unfolds a secret that can only be known through eq(3.9).

From eq(3.4) we can also easily show that $\vec k . \vec E = \vec k .
\vec B = 0$, i.e., PWS of ME are {\it transverse} waves.

We can rewrite eq(3.4) as
\bb
k\ga_0 \ga_0 F \ga_0 = 0
\ee
and since $k \ga_0 = k_0 + \vec k, \ \ga_0 F \ga_0 = - \vec E + \ia
\vec B$ we have
\bb
\vec k f = k_0 f .
\ee

Now, we recall that in $\clif^+(M)$
(where the typical fiber is isomorphic to the
Pauli algebra $\clif_{3,0}$) we can introduce the operator of space
conjugation denoted by $*$ \cite{REF-6} such that writing $f = \vec e +
\ia \vec
b$ we have
\bb
f^* = - f \ \ ; \ \ k^*_0 = k_0 \ \ ; \ \ \vec k^* = - \vec k .
\ee
We can now interpret the two solutions of $k^2 = 0$, i.e.,
$k_0=|\vec k|$ and $k_0 = -|\vec k|$ as corresponding to the
solutions $k_0 f = \vec kf$ and $k_0 f^* = - \vec kf^*$; $f$ and
$f^*$ correspond in quantum theory to ``photons" which are of
positive or negative helicities. We can interpret $k_0 = |\vec k|$ as
a particle and $k_0 = -|\vec k|$ as an antiparticle.

Summarizing we have the following  important facts concerning PWS of
ME: (i) the propagation vector is light-like, $k^2=0$; (ii) the
field invariants are null, $F^2=0$; (iii) the PWS are transverse
waves, i.e., $\vec k . \vec E = \vec k . \vec B = 0$.

\subsection{Subluminal Solutions of Maxwell Equations and Purely
Electromagnetic Particles (PEPs)}

In order to present the subluminal and superluminals solutions of
Maxwell equations we need the following result \cite{rodvaz95}:

Let $A \in \sec
\bigwedge^1(M) \subset \sec \clif (M)$ be the
vector potential. We
fix the Lorentz gauge, i.e., $\dirac \cdot A = - \delta A =0$ such that
$F=\dirac A =
(d-\delta)A = dA$. We have the following theorem: \\

\noindent {\bf Theorem:} Let $\pi \in \sec \bigwedge^2(M)
\subset \sec \clif(M)$ be the
so called Hertz potential. If $\pi$
satisfies the wave equation, i.e, $\partial^2 \pi =
\eta^{\mu\nu}\partial_\mu\partial_\nu \pi = -(d \delta + \delta d)\pi
=0$, and if $A =
-\delta\pi$, then $F = \dirac A$
satisfies the
Maxwell equations $\dirac F =0$.

The proof is trivial. Indeed $A = - \delta \pi$, then $\delta A = -
\delta^2
\pi = 0$ and $F =\dirac A =dA$. Now $\dirac F = (d-\delta) \
(d-\delta)A = -
(d\delta + \delta d) A = \delta d(\delta \pi) = - \delta^2 d \pi = 0$
since $\delta d \pi = -d \delta \pi$ from $\partial^2 \pi = 0$.

We take $\Phi \in \sec (\bwe^0(M) \oplus \bwe^4(M)) \sub \sec
\clif(M)$ and consider \cite{vazrod95} the following Hertz potential
$\pi \in \sec
\bwe^2(M) \sub \sec \clif(M)$:
\bb
\pi = \Phi \ga^1 \ga^2 .
\ee
We now write

\bb
\Phi(t, \vec x) = \phi(\vec x) e^{\ga_5 \Om t} .
\ee
Since $\pi$ satisfies the wave equation, we have

\bb
\nab^2 \phi(\vec x) + \Om^2 \phi(\vec x) = 0 \, .
\ee

Solutions of eq(3.15) (the Helmholtz equation) are well known. Here,
we consider the simplest solution in spherical coordinates,

\bb
\phi(\vec x) = C \frac{\sin\Om r}{r} \ \ , \ \ r =
\sqrt{x^2+y^2+z^2},
\ee
where $C$ is an arbitrary real constant.We obtain the  following
stationary electromagnetic field, which
is at rest in the reference frame $I$ where $\lan x^\mu\ran$ are
naturally adapted coordinates:
\bega
F_0 & = & \frac{C}{r^3} [\sin \Om t(\al\Om r\sin\theta \sin\rho - \be
\cos
\theta \sin\theta \cos\rho) \ga^0 \ga^1 \nb  \\
&& - \sin \Om t (\al\Om r\sin\theta \cos\rho + \be \sin \theta
\cos\theta \sin\rho)
\ga^0 \ga^2 \nb  \\
&& + \sin \Om t(\be\sin^2 \theta - 2\al)\ga^0 \ga^3 + \cos\Om t
(\be\sin^2 \theta -  2\al)  \ga^1 \ga^2 \\
&& + \cos\Om t(\be\sin\theta \cos\theta \sin\rho + \al\Om r \sin \theta
\cos\rho)\ga^1 \ga^3 \nb \\
&& + \cos\Om t(-\be\sin\theta \cos\theta \cos\rho + \al\Om \sin \theta
\sin\rho)\ga^2 \ga^3 ] \nb
\ega
with $\al=\Om r \cos\Om r - \sin\Om r$ and $\be = 3\al + \Om^2 r^2
\sin \Om r$. Observe that $F_0$ is regular at the origin and vanishes
at infinity. Let us rewrite the solution using the Pauli-algebra in
$\clif^+(M)$. Writing $(\ia \equiv \ga_5)$

\bb
F_0 = \vec E_0 + \ia \vec B_0
\ee
we get

\bb
\vec E_0 = - \vec W \sin \Om t \ \ , \ \ \vec B_0=\vec W \cos\Om t
\ee
with

\bb
\vec W = C \left( \frac{\al\Om y}{r^3} - \frac{\be x z}{r^5} , -
\frac{\al\Om x}{r^3} - \frac{\be y z}{r^5}, \frac{\be(x^2+y^2)}{r^5}
- \frac{2\al}{r^3}\right) .
\ee
We verify that $\divi \vec W = 0, \ \divi \vec E_0 = \divi \vec B_0 =
0, \rot \vec E_0 + \pa \vec B_0/\pa t = 0, \rot \vec B_0 - \pa \vec
E_0/\pa t = 0$, and

\bb
\rot \vec W = - \Om \vec W.
\ee

Now, it is well known that $T_0 = {\dis\frac{1}2} \wid F \ga_0
F$ is the 1-form representing the energy density and the Poynting
vector \cite{REF-5,REF-6} . It follows that $\vec E_0 \times \vec B_0 =
0$, i.e., the
solution has zero angular momentum. The energy density
$u = S^{00}$ is given by

\bb
u = \frac{1}{r^6} [\sin^2 \theta (\Om^2 r^2 \al^2 + \cos^2 \theta
\be^2) +
(\be\sin^2 \theta - 2\al)^2]
\ee
Then $\int\!\int\!\int_{\R^3} u \, d\ven = \infi$. In \cite{upw}
it is discussed how to generate  finite
energy solutions. It can be constructed by considering ``wave packets"
with a distribution of intrinsic frequencies $F(\Om)$ satisfying
appropriate conditions. Many possibilities exist, but they will not be
discussed here. Instead, we prefer to direct our attention to
eq.(3.21). As it is well known, this is a very important equation
(called the force free equation \cite{wai95}) that appears e.g. in
hydrodynamics and in several different situations in plasma
physics \cite{ree94}. The following considerations are more important.

Einstein, among others, (see \cite{wai95} for a review) studied
the possibility of constructing PEPs. He started from Maxwell
equations for a PEP configuration described by an electromagnetic
field $F_p$ and a current density $J_p = \rho_p \gamma_0 + j_p^i
\gamma_i$, where
\bb
\dirac F_p = J_p
\ee
and rightly concluded that the condition for existence of PEPs is
\bb
J_p . F_p = 0.
\ee
This condition implies, in vector notation,
\bb
\rho_p \vec E_p = 0 \ \ , \ \ \vec j_p . \vec E_p = 0 \ \ , \ \ \vec
j_p \times \vec B_p = 0 \, .
\ee

From eq.(3.25) Einstein concluded that the only possible solution of
eq.(3.23) with the subsidiary condition given by eq.(3.24) is $J_p =
0$. However, this conclusion is correct  only
if $J^2_p > 0$, i.e., if $J_p$ is a time-like current density. However,
if we suppose that $J_p$ can be spacelike, i.e., $J^2_p < 0$, there
exists a reference frame where $\rho_p = 0$ and a possible solution of
eq.(3.24) is
\bb
\rho_p = 0 \ \ , \ \ \vec E_p . \vec B_p = 0 \ \ , \ \ \vec j_p =
kC\vec B_p ,
\ee
where $k = \pm 1$ is called the chirality of the solution and $C$ is
a real constant. In \cite{wai95,waietal96} static solutions of eqs.
(3.23) and (3.24) are
exhibited where $\vec E_p = 0$. In this case we can verify that $\vec
B_p$ satisfies.
\bb
\nab \times \vec B_p = kC\vec B_p .
\ee
Now, if we choose $F \in \sec \bwe^2 (M) \sub \sec \clif(M)$ such
that
\bb
\begin{array}{c}
F_0 = \vec E_0 + \ia \vec B_0 \, , \\
\vec E_0 = - \vec B_p \cos \Om t \ \ , \ \ \vec B_0 = \vec B_p \sin
\Om t \, ,
\end{array}
\ee
we immediately realize that
\bb
\dirac F_0 = 0 \, .
\ee
This is an amazing result, since it means that the free Maxwell
equations may have stationary solutions that model PEPs. In such
solutions the structure of the field $F_0$ is such that we can write,
e.g.,

\bb
\begin{array}{c}
F_0 = F_{p}^{'} + \ov F = \mbox{\bf i} \vec{W} \cos \omega t - \vec{W}
\sin \Omega t \, , \\
\dirac F_{p}^{'} = -\dirac \ov F = J_{p}^{'} \, ,
\end{array}
\ee
i.e., $\dirac F_0=0$ is equivalent to a field plus a current. This
opens several
interesting possibilities for modelling PEPs (see also
\cite{wai95,waietal96}) and we
discuss more this issue in another publication.

We observe that moving subluminal solutions of ME can be easily
obtained
choosing as Hertz potential, e.g.,
\bega
&& \h \pi^<(t,\vec x) = C \frac{\sin\Om\xi_<}{\xi_<} \exp [\ga_s(\om_<
t-k_< z)]\ga_1 \ga_2 \\
&& \h\h \om^2_< - k^2_< = \Om^{2}_{<} \nb \\
&& \xi_< = [x^2+y^2+ \ga^2_< (z-vt)^2] \\
&& \h \ga_< = \frac{1}{\sqrt{1-v^2_<}} \ , \ \ v_< = d\om_</ dk_< \nb
\ega
We are not going to write explicitly the expression for $F^<$
corresponding to $\pi^<$ because it is very long and will not be used
in what follows.

We end this section with the following observations: (i) In general
for subluminal solutions of ME (SSME) the propagation vector
satisfies an equation like eq(3.32).  (ii) As can be easily verified,
for a SSME the field invariants are non-null. (iii) A SSME is not a
tranverse wave. This can be seen explicitly from eqs(3.19) and (3.20)
and a special Lorentz tranformation with parameter $v_<$.

Conditions (i), (ii), (iii) are in contrast with the case of the PWS of
ME. In \cite{vazrod93,vazrod95} it is shown that for free
electromagnetic
fields $(\dirac F=0)$ such that $F^2 \neq 0$, there exists a
Dirac-Hestenes
equation
 for $\psi \in \sec (\bwe^0(M) + \bwe^2(M) + \bwe^4(M))
\sub \sec \clif(M)$ where $F = \psi \ga_1 \ga_2 \wid \psi$. This was
 the reason why \cite{vazrod95} discovered subluminal and
superluminal solutions of Maxwell equations (and also of Weyl
equations) which solve the Dirac-Hestenes equation.

\subsection{The Superluminal Electromagnetic $X$-Wave (SEXW)}

To simplify the matter in what follows we now suppose that the
functions
$\Phi_{X_n}$  and $\Phi_{XBB_n}$ below, which are
superluminal solutions of the scalar wave equation
\cite{lugreen92,upw}, are 0-forms
sections of the complexified Clifford bundle $\clif_C(M) = \C \ot
\clif(M)$. Sections of $\clif_C(M)$ are like in eq.(\ref{2e1}) with the
coefficients $s, \, v_\mu \, , b_{\mu\nu} \ldots$ being complex
functions. We have

\bb
\Phi_{X_n} (t, \vec x) = e^{in\theta} \int^\infi_0 B(\ov k) J_n (\ov k
\rho\sin\eta) e^{-\ov k[a_0 - i(z \cos\eta-t)]} d\ov k
\ee
where $n=0,\, 1, \, 2, \ldots$ and $\eta$ is a constant, called the
axicon angle.
Choosing $B(\ov k) = a_0$, we have
\bega
&& \h \Phi_{XBB_n} (t,\vec x) = \frac{a_0(\rho\sin \eta)^n
e^{in\theta}}{\sqrt{M}(\tau+\sqrt{M})^n} \\
&& M = (\rho\sin\eta)^2 + \tau^2 \ \ \ ; \ \ \  \tau = [a_0 -
i(z\cos\eta-t)].
\ega
 Further, we suppose now that the Hertz potential
$\pi$, the vector potential A and the corresponding electromagnetic
field $F$ are appropriate sections of $\clif_C(M)$. We
take
\bb \label{3e36}
\pi=\Phi \ga_1 \ga_2 \in \sec \C \ot \bwe^2(M) \sub \sec
\clif_C(M),
\ee
where $\Phi$ can be $\Phi_{X_n}$ or $\Phi_{XBB_n}$.  Let us start by
giving the
explicit form of the $F_{XBB_n}$ i.e., the SEXWs. In this case writing
$\vec{\pi} = \vec{\pi}_e + \mbox{\bf i} \vec{\pi}_m$ eq.(\ref{3e36})
gives $\pi=\vec \pi_m$ and

\bb
\vec \pi_m = \Phi_{XBB_n} \mbox{\boldmath$z$} \, ,
\ee
where \mbox{\boldmath$z$}   is the versor of the $z$-axis. Also, let
$\ro,
\thetao$ be respectively  the versors of the $\rho$ and $\theta$
directions
where $(\rho, \theta, z)$ are the usual cylindrical coordinates.

Writing
\bb
F_{XBB_n} = \vec E_{XBB_n} + \ga_5 \vec B_{XBB_n}
\ee
we obtain

\begin{equation}
\vec E_{XBB_n} = -\frac{\ro}{\rho} \frac{\pa^2}{\pa t\pa\theta}
\Phi_{XBB_n} + \thetao \frac{\pa^2}{\pa t\pa \rho} \Phi_{XBB_n}
\end{equation}
\begin{equation}
\vec B_{XBB_n} =\ro \frac{\pa^2}{\pa \rho\pa z}
\Phi_{XBB_n} + \thetao \frac{1}{\rho} \frac{\pa^2}{\pa \theta \pa z}
\Phi_{XBB_n} +
\zo \left(\frac{\pa^2}{\pa z^2} \Phi_{XBB_n} - \frac{\pa^2}{\pa
t^2} \Phi_{XBB_n} \right)
\end{equation}

Explicitly we get for the components in cylindrical coordinates,

\h $(\vec E_{XBB_n})_\rho = -{\displaystyle \frac{1}{\rho}
n\frac{M_3}{\sqrt{M}}}
\Phi_{XBB_n}$ \hfill (3.41a)

\h $(\vec E_{XBB_n})_\theta = {\dis\frac{1}{\rho} i\frac{M_6}{\sqrt{M}
M_2}}
\Phi_{XBB_n}$ \hfill (3.41b)

\h $(\vec B_{XBB_n})_\rho = \cos\eta (\vec E_{XBB_n})_\theta$ \hfill
(3.41c)

\h $(\vec B_{XBB_n})_\theta = -\cos\eta (\vec E_{XBB_n})_\rho$ \hfill
(3.41d)

\h $(\vec B_{XBB_n})_z = - \sin^2\eta {\dis\frac{M_7}{\sqrt{M}}}
\Phi_{XBB_n}$. \hfill
(3.41e)

The functions $M_i (i=2,\ld,7)$ in (3.41) are

\h $M_2 = \tau + \sqrt{M}$ \hfill (3.42a)

\h $M_3 = n + {\dis\frac{1}{\sqrt{M}}} \tau$ \hfill (3.42b)

\h $M_4 = 2n + {\dis\frac{3}{\sqrt{M}}}\tau$ \hfill (3.42c)

\h $M_5 = \tau + n\sqrt{M}$  \hfill (3.42d)

\h $M_6 = (\rho^2 \sin^2\eta {\dis\frac{M_4}{M}} - n M_3)M_2 + n \rho^2
\sin^2 \rho {\dis\frac{M_5}{M}}$  \hfill (3.42e)

\h $M_7 = (n^2-1) {\displaystyle \frac{1}{\sqrt{M}} + 3 n \frac{1}{M}}
\tau +
3  {\dis\frac{1}{\sqrt{M^3}}}\tau^2$  \hfill (3.42f)

We immediately see from eq(3.41) that the $F_{XBB_n}$ are indeed
superluminal UPWs solutions of ME, propagating with speed $1/\cos\eta$
in the $z$-direction. That $F_{XBB_n}$ are undistorted progressive
waves is trivial and that
they propagate with speed $c_1=1/\cos\eta$ follows because
$F_{XBB_n}$ depends only on the combination of variables $(z-c_1t)$
and any  derivatives of $\Phi_{XBB_n}$ will keep the
$(z-c_1t)$ dependence structure. More details can be found in
\cite{upw} where we show how to generate finite aperture approximations
to the SEXWs. In particular, the name $X$-wave is due to the shape of
these waves.

\section{The Equivalence Between Maxwell and Dirac Equations}

Let us consider the generalized Maxwell equations

\begin{equation} \label{4e1}
\dirac F = {\cal J} \, ,
\end{equation}
where $\dirac = \gamma^\mu \partial_\mu$ is the Dirac operator and
${\cal J}$ is the elctromagnetic current (an electric current $J_e$
plus a magnetic monopole current $-\gamma_5 J_m$, where $J_e$, $J_m \in
\sec \bwe^1M \subset \clif (M)$). In \cite{vazrod93,vazrod95} we proved
that $F$ can be written as

\begin{equation} \label{4e2}
F = \psi \gamma_{21} \tilde{\psi} \, ,
\end{equation}
where $\psi \in \sec \clif^+(M)$ is a Dirac-Hestenes spinor field. If
we use eq.(\ref{4e2}) in eq.(\ref{4e1}) we get


\begin{equation}
\label{eq.41}
\dirac(\psi\gamma_{21}\tilde{\psi}) = \gamma^{\mu}\partial_{\mu}
(\psi\gamma_{21}\tilde{\psi}) =
\gamma^{\mu}(\partial_{\mu}\psi
\gamma_{21}\tilde{\psi} + \psi \gamma_{21}\partial_{\mu}\tilde{\psi}) =
{\cal J}  .
\end{equation}
But $\psi\gamma_{21}\partial_\mu \tilde{\psi} = -(\partial_{\mu}\psi
\gamma_{21}\tilde{\psi})\, \tilde{}$, and since reversion does not
change the sign of scalars and of pseudo-scalars (4-vectors), we have
that
\begin{equation}
\label{eq.42}
2\gamma^{\mu}\langle\partial_{\mu}\psi\gamma_{21}\tilde{\psi}\rangle_2
= {\cal J} .
\end{equation}
There is a more convenient way of rewriting the above equation. Note
that
\begin{equation}
\label{eq.43}
\gamma^{\mu}\langle\partial_{\mu}\psi\gamma_{21}\tilde{\psi}\rangle_2 =
\dirac \psi\gamma_{21}\tilde{\psi} -
\gamma^{\mu}\langle\partial_{\mu}\psi\gamma_{21}\tilde{\psi}\rangle_0
-
\gamma^{\mu}\langle\partial_{\mu}\psi\gamma_{21}\tilde{\psi}\rangle_4 ,
\end{equation}
and if we define the vectors
\begin{equation}
\label{eq.44}
j =
\gamma^{\mu}\langle\partial_{\mu}\psi\gamma_{21}\tilde{\psi}\rangle_0 ,
\end{equation}
\begin{equation}
\label{eq.45}
g = \gamma^{\mu}\langle\partial_{\mu}\psi\gamma_5
\gamma_{21}\tilde{\psi}\rangle_0  ,
\end{equation}
we can rewrite eq.(\ref{eq.42}) as
\begin{equation}
\label{eq.46}
\dirac \psi\gamma_{21}\tilde{\psi} =
\left[ \frac{1}{2}{\cal J} + \left( j + \gamma_5 g \right) \right]  .
\end{equation}

Eq.(\ref{eq.46}) is a spinorial representation of Maxwell equations.
 In the case where $\psi$ is non-singular (which
corresponds to non-null electromagnetic fields) we have

\begin{equation}
\label{eq.47}
\dirac\psi\gamma_{21} = \frac{{\rm e}^{\gamma_5 \beta}}{\rho}
\left[ \frac{1}{2}{\cal J} +
\left( j + \gamma_5 g \right) \right] \psi  .
\end{equation}
Eq.(\ref{eq.47}) has been proved \cite{vazrod93} to be equivalent to
the spinorial representation of Maxwell equations obtained originally
by  \cite{cam90} in terms of the
usual covariant Dirac spinor field.

The spinorial equation (\ref{eq.47}) representing Maxwell equations,
written in that form, does not appear to have any relationship with
Dirac-Hestenes equation \ref{2e7}. However, we shall make some
modifications
on it in such a way as to put it in a form that suggests a very
interesting
and intriguing relationship between them, and consequently between
electromagnetism and quantum mechanics.

Since $\psi$ is supposed to be non-singular ($F$ non-null) we can use
the canonical decomposition  of $\psi$ and write $\psi = \rho e^{\beta
\gamma_5 /2} R$, with $\rho$, $\beta \in \sec \bwe^0 M \subset \sec
\clif
(M)$ and $R \in \mbox{Spin$_+(1,3)$}\  \forall x \in M$. Then

\begin{equation}
\label{eq.48}
\partial_{\mu}\psi = \frac{1}{2}\left( \partial_{\mu}\ln{\rho} +
\gamma_5 \partial_{\mu}\beta + \Omega_{\mu} \right)\psi ,
\end{equation}
where we defined
\begin{equation}
\label{eq.49}
\Omega_{\mu} = 2(\partial_{\mu}R)\tilde{R}  .
\end{equation}

Using this expression for $\partial_{\mu}\psi$ into the definitions
of the vectors $j$ and $g$ (eqs.(\ref{eq.44},\ref{eq.45})) we obtain
that
\begin{equation}
\label{eq.50}
j = \gamma^{\mu}(\Omega_{\mu}\cdot S)\rho\cos{\beta} +
\gamma_{\mu}[\Omega_{\mu}\cdot(\gamma_5 S)]\rho\sin{\beta} ,
\end{equation}
\begin{equation}
\label{eq.51}
g = \gamma^{\mu}[(\Omega_{\mu}\cdot (\gamma_5 S)]\rho\cos{\beta} -
\gamma_{\mu}(\Omega_{\mu}\cdot S)\rho\sin{\beta} ,
\end{equation}
where we defined the bivector $S$ by
\begin{equation}
\label{eq.52}
S = \frac{1}{2}\psi\gamma_{21}\psi^{-1} =
\frac{1}{2}R\gamma_{21}\tilde{R}  .
\end{equation}
A more convenient expression can be written. Let $v$ be given by
$\rho v = J = \psi \gamma_0 \tilde{\psi}$, and $v_{\mu} =
v\cdot\gamma_{\mu}$.
Define the bivector $\Omega = v^{\mu}\Omega_{\mu}$ and the scalars
$\Lambda$ and $K$ by
\begin{equation}
\label{eq.53}
\Lambda = \Omega \cdot S ,
\end{equation}
\begin{equation}
\label{eq.54}
K = \Omega \cdot (\gamma_5 S) .
\end{equation}
Using these definitions we have that
\begin{equation}
\label{eq.55}
\Omega_{\mu}\cdot S = \Lambda v_{\mu} ,
\end{equation}
\begin{equation}
\label{eq.56}
\Omega_{\mu}\cdot(\gamma_5 S) = K v_{\mu}  ,
\end{equation}
and for the vectors $j$ and $g$:
\begin{equation}
\label{eq.57}
j = \Lambda v \rho\cos{\beta} + K v \rho\sin{\beta} = \lambda \rho v ,
\end{equation}
\begin{equation}
\label{eq.58}
g = K v \rho\cos{\beta} - \Lambda v \rho \sin{\beta} = \kappa \rho v ,
\end{equation}
where we defined
\begin{equation}
\label{eq.59}
\lambda = \Lambda \cos{\beta} + K \sin{\beta} ,
\end{equation}
\begin{equation}
\label{eq.60}
\kappa = K \cos{\beta} - \Lambda \sin{\beta}  .
\end{equation}
The spinorial representation of Maxwell equations is written
now as
\begin{equation}
\label{eq.61}
\dirac \psi\gamma_{21} = \frac{{\rm e}^{\gamma_5 \beta}}{2\rho}{\cal
J}\psi +
\lambda\psi\gamma_0 + \gamma_5 \kappa\psi\gamma_0  .
\end{equation}
If ${\cal J} = 0$ (free case) we have that
\begin{equation}
\label{eq.62}
\dirac \psi\gamma_{21} =
\lambda\psi\gamma_0 + \gamma_5 \kappa\psi\gamma_0  ,
\end{equation}
which is very similar to the Dirac-Hestenes equation (\ref{2e7}).

In order to go a step further into the relationship between those
equations, we remember that the electromagnetic field has six
degrees of freedom, while a Dirac-Hestenes spinor field has
eight degrees of freedom; we are free therefore to impose two
constraints on $\psi$ if it is to represent an electromagnetic
field. We choose these two constraints as
\begin{equation}
\label{eq.63}
\dirac \cdot j = 0 \, \, \, \, \, \, {\rm and} \, \, \, \, \, \,
\dirac \cdot g = 0 .
\end{equation}
Using eqs.(\ref{eq.57},\ref{eq.58}) these two constraints become\
\begin{equation}
\label{eq.64}
\dirac \cdot j = \rho \dot{\lambda} + \lambda\dirac\cdot J = 0 ,
\end{equation}
\begin{equation}
\label{eq.65}
\dirac \cdot g = \rho \dot{\kappa} + \kappa\dirac\cdot J = 0 ,
\end{equation}
where $J = \rho v$ and $\dot{\lambda} = (v\cdot\dirac)\lambda$,
$\dot{\kappa} = (v\cdot\dirac)\kappa$. These conditions imply that
\begin{equation}
\label{eq.66}
\kappa\dot{\lambda} = \lambda\dot{\kappa}  ,
\end{equation}
which gives ($\lambda \neq 0$):
\begin{equation}
\label{eq.67}
\frac{\kappa}{\lambda} = {\rm const.} = -\tan{\beta_0} ,
\end{equation}
or from eqs.(\ref{eq.59},\ref{eq.60}):
\begin{equation}
\label{eq.68}
\frac{K}{\Lambda} = \tan{(\beta - \beta_0)} .
\end{equation}

Now we observe that $\beta$ is the angle of the duality rotation
from $F$ to $F^{\prime} = {\rm e}^{\gamma_5 \beta}F$. If we perform
another duality rotation by $\beta_0$ we have $F \mapsto {\rm
e}^{\gamma_5
(\beta + \beta_0)}F$, and for the Yvon-Takabayasi angle $\beta \mapsto
\beta + \beta_0$. If we work therefore with an electromagnetic field
duality rotated by an additional angle $\beta_0$, the above
relationship becomes
\begin{equation}
\label{eq.69}
\frac{K}{\Lambda} = \tan{\beta} .
\end{equation}
This is, of course, just a way to say that we can choose the constant
$\beta_0$ in eq.(\ref{eq.67}) to be zero. Now,
this expression gives
\begin{equation}
\label{eq.70}
\lambda = \Lambda \cos{\beta} + \Lambda \tan{\beta}\sin{\beta} =
\frac{\Lambda}{\cos{\beta}}  ,
\end{equation}
\begin{equation}
\label{eq.71}
\kappa = \Lambda \tan{\beta}\cos{\beta} - \Lambda \sin{\beta} = 0 ,
\end{equation}
and the spinorial representation (\ref{eq.62}) of the free Maxwell
equations becomes
\begin{equation}
\label{eq.72}
\dirac\psi\gamma_{21} = \lambda\psi\gamma_0  .
\end{equation}

Note that $\lambda$ is such that
\begin{equation}
\label{eq.73}
\rho\dot{\lambda}  = -\lambda \dirac\cdot J .
\end{equation}
The current $J = \psi\gamma_0\tilde{\psi}$ is not conserved unless
$\lambda$ is constant. If we suppose also that
\begin{equation}
\label{eq.74}
\dirac \cdot J = 0
\end{equation}
we must have
\begin{equation}
\label{eq.75}
\lambda = {\rm const.}
\end{equation}

Now, throughout these calculations we have assumed $\hbar = c = 1$.
We observe that in eq.(\ref{eq.72}) $\lambda$ has the units of
(length)$^{-1}$, and if we introduce the constants $\hbar$ and $c$
we have to introduce another constant with unit of mass. If we
denote this constant by $m$ such that
\begin{equation}
\label{eq.76}
\lambda = \frac{mc}{\hbar} ,
\end{equation}
then eq.(\ref{eq.72}) assumes a form which is identical to
Dirac equation:
\begin{equation}
\label{eq.77}
\partial\psi\gamma_{21} = \frac{mc}{\hbar}\psi\gamma_0  .
\end{equation}

It is true that we didn't prove that eq.(\ref{eq.77}) is really
Dirac equation since the constant $m$ has to be identified in
this case with the electron's mass. However in \cite{vazrod95} we
present several arguments based on the stochastic interpretation of
quantum mechanics which suggest that we must take the identification
seriously.


\section{Some Properties of the Linear and Non-Linear Dirac-Hestenes
Equation}

\subsection{Projection Operators and Energy-Momentum Tensors}

We recall that the free DHE

\begin{equation} \label{5e1}
\dirac \psi \gamma_{12} + m \psi \gamma_0 = 0
\end{equation}
has the following plane wave solutions:

\begin{equation}
\begin{array}{rcl}
\psi^{(+)}_{\uparrow} & = & \sqrt{\rho} e^{-\gamma_{21} m t} \, ;  \\
\psi^{(+)}_{\downarrow} & = &  \sqrt{\rho} \gamma_{31} e^{-\gamma_{21}
m t} \, ; \\
\psi^{(-)}_{\uparrow} & = & \sqrt{\rho} \gamma_5 (\gamma_{12}
e^{\gamma_{21} m t } ) \, ; \\
\psi^{(-)}_{\downarrow} & = & \sqrt{\rho} \gamma_5 \gamma_{31}
(\gamma_{12} e^{\gamma_{21} m t} )\, .
\end{array}
\end{equation}
Observe that for the solutions $(+)$ we have $\beta = 0$, and for the
solutions $(-)$ we have $\beta = \pi$.

The solutions $(+)$ [$(-)$] are solutions of positive [negative] energy
and the arrows indicate spin-up ($\uparrow$) and spin-down
($\downarrow$). To see this we consider the energy projection operators
$\proje_{\pm} \cite{REF-4,lo94}$:

\begin{equation} \label{5e3}
\proje_{\pm} (\psi) = \frac{1}{2} [\psi \pm \gamma_0 \psi \gamma_0 ]
\end{equation}
and the spin projector operators $\prospin_{\pm}$:

\begin{equation} \label{5e4}
\prospin_{\pm} (\psi) = \frac{1}{2} [\psi \pm \gamma_{21} \psi
\gamma_{21} ] \, .
\end{equation}

It is easy to see that $\proje_{\pm}^2 = \proje_{\pm}$, $\proje_{\mp}
\proje_{\pm} = 0$, $\proje_{+} + \proje_{-} = 1$ and $\prospin_{\pm}^2
= \prospin_{\pm}$, $\prospin_{\mp} \prospin_{\pm} = 0$, $\prospin_{+} +
\prospin_{-} = 1$.   Then,

\begin{equation}
\begin{array}{c}
\proje_{+} \prospin_{+} \psi_{\uparrow}^{(+)} = \prospin_{+} \proje_{+}
\psi_{\uparrow}^{(+)} = \psi_{\uparrow}^{(+)}  \, ; \\
\proje_{+} \prospin_{-} \psi_{\downarrow}^{(+)} = \prospin_{-}
\proje_{+} \psi_{\downarrow}^{(+)} = \psi_{\downarrow}^{(+)} \, ;  \\
\proje_{-} \prospin_{+} \psi_{\uparrow}^{(-)} = \prospin_{+} \proje_{-}
\psi_{\uparrow}^{(-)} = \psi_{\uparrow}^{(-)} \, ;  \\
\proje_{-} \prospin_{-} \psi_{\downarrow}^{(-)} = \prospin_{-}
\proje_{-} \psi_{\downarrow}^{(-)} = \psi_{\downarrow}^{(-)} \, .
\end{array}
\end{equation}
The fact that the Takabayasi angle $\beta$ is $0$ or $\pi$ for these
solutions is very interesting and a mistery. We already saw that the
Takabayasi angle appears in the NLDHE, which is equivalent to Maxwell
equation, and we are going to disclose some of the secrets of the
variable $\beta$ in what follows.

Before going on we recall that  the solutions of the Dirac equation
(coupled with the electromagnetic field) for the hydrogen atom are
solutions with variable $\beta$, i.e. $\beta = \beta (x)$
\cite{quili71}. But it is very interesting that \cite{kru91} obtained
solutions for the hydrogen atom with $\beta =0$ or $\pi$.

We return now to our NLDHE  (eq.(\ref{eq.47})) which we can also write
as

\begin{equation} \label{5e6}
\dirac \psi \gamma_{21} = \Lambda \psi \gamma_0 e^{ \beta \gamma_5} +
\gamma_5 K \psi \gamma_0 e^{\beta \gamma_5} + \frac{1}{2\rho} e^{\beta
\gamma_5} {\cal J} \psi \, .
\end{equation}
For ${\cal J}=0$ and $K=0$ we have

\begin{equation} \label{5e7}
\dirac \psi \gamma_{21} = \Lambda \psi \gamma_0 e^{\beta \gamma_5} \, ,
\end{equation}
which has been studied extensively by \cite{daviau93}, but in a context
different from the present one. Note that this result is consistent
with eq.(\ref{eq.77}) if $\beta =0$ or $\beta =\pi$.

Equation (\ref{5e7}) is non linear. Except for the term $e^{\beta
\gamma_5}$, which introduces the non-linearity, this equation is like
the DHE if $\Lambda=m$. We  briefly discuss, before arriving at the
main topics of this section, the fact that eq.(\ref{5e7}) is more
satisfactory than the DHE.

To start with, we know \cite{hest73} that the energy-momentum operator
in the Clifford bundle formalism is given by

\begin{equation}
\hat{p} \psi = \dirac \psi \gamma_{21} \, .
\end{equation}
Since $\gamma_0 \dirac = \partial_0 + \nabla$, $\gamma_0 p = p_0 -
\vec{p}$, we have

\begin{equation}
\hat{E} \psi = i \partial_0 \psi \, , \quad \hat{\vec{p}} \psi = -i
\nabla \psi \; ,
\end{equation}
which are the usual definitions of the operators $\hat{E}$ and
$\hat{\vec{p}}$. We can then write eq.(\ref{5e7}) as

\begin{equation} \label{5e10}
\hat{p} \psi = m c \psi  \gamma_0 e^{\beta \gamma_5}
\end{equation}
and if $\psi$ is a nonsigular eigenvector of $\hat{p}$ we have, after
multiplying eq.\ref{5e10} by $\psi^{-1}$, that

\begin{equation}
p=mv \, ,
\end{equation}
with $\displaystyle v = \frac{1}{\rho} \psi \gamma_0 \tilde{\psi}$.

On the other hand, writing the DHE (eq.(\ref{5e1})) as

\begin{equation}
\hat{p} \psi = m \psi \gamma_0
\end{equation}
we get for the nonsingular eigenvectors of $\hat{p}$ that

\begin{equation} \label{5e13}
p = e^{\beta \gamma_5} mv \, .
\end{equation}
Since $p$ and $v$ are 1-forms, eq.(\ref{5e13}) implies that $\beta = 0$
or $\beta = \pi$, and we get

\begin{equation}  \label{5e14}
p = \pm mv \, ,
\end{equation}
i.e., the case $\beta = \pi$ introduces a negative sign without any
obvious physical meaning. Equation (\ref{5e14}) implies that in order
to make sense the DHE for a free particle must have $\beta = 0$ or
$\beta = \pi$, and we have the problem of interpreting the negative
energy. The known solution is to say that such solutions correspond to
antiparticles.

Now,  the NLDHE (\ref{5e7}) does not present any problem, because
$p=mv$ for whatever angle $\beta$. The following point concerning
eq.(\ref{5e7}) is important.
The superposition principle of quantum mechanics remains valid if we
only superpose solutions with the same $\beta$. This means that fixed
values of $\beta$ imply in a superselection rule. To understand this
fact we write $\psi = \phi e^{\beta \gamma_5} $ and supposing $\beta =
\mbox{constant}$ eq.(\ref{5e7}) becomes

\begin{equation}  \label{5e15}
\dirac \phi \gamma_{21} = m \phi \gamma_{0} \, ,
\end{equation}
which is the DHE for $\phi = \sqrt{\rho}R$. Then, the solutions of
eq.(\ref{5e15}) for $\phi$ are identical to the solutions of the DHE
for $\beta = 0$. We have the solutions

\begin{equation} \label{5e16}
\psi_{\uparrow} = \sqrt{\rho} e^{\beta \gamma_5 /2} e^{-\gamma_{21}mt}
\, ;
\end{equation}
\begin{equation}  \label{5e17}
\psi_{\downarrow} = \sqrt{\rho}  e^{\beta\gamma_5/2}  \gamma_{31}
e^{-\gamma_{21}mt} \, ;
\end{equation}
where $\beta$ is arbitrary but constant. Since $\gamma_5$ commutes with
$\gamma_{21}$, it is obvious that these solutions are eigenvectors of
the spin projector operators $\prospin_{\pm}$ (eq.(\ref{5e4})). On the
other hand, equations (\ref{5e16}) and (\ref{5e17})  are not
eigenvectors of the energy projector operators $\proje_{\pm}$
(eq.(\ref{5e3})). Indeed, we have

\begin{equation}
\proje_{+} (\psi_{\uparrow\downarrow}) = \left[ \cos^2 \frac{\beta}{2}
- \gamma_5 \frac{1}{2} \sin \beta \right] \psi_{\uparrow\downarrow} \;
;
\end{equation}
\begin{equation}
\proje_{-} (\psi_{\uparrow\downarrow}) = \left[ \gamma_5 \frac{1}{2}
\sin \beta +  \sin^2 \frac{\beta}{2} \right] \psi_{\uparrow\downarrow}
\; .
\end{equation}

It follows that $\psi_{\uparrow\downarrow}$ are eigenvectors of
$\proje_{\pm}$ only for $\beta =0$ or $\beta =\pi$ and in these cases
we have

\begin{equation}
\proje_{+} (\psi_{\uparrow\downarrow}) =
\left\{
\begin{array}{cl}
\psi_{\uparrow\downarrow} & (\beta = 0) \\
0 &  (\beta = \pi)
\end{array}
\right.
\end{equation}
\begin{equation}
\proje_{-} (\psi_{\uparrow\downarrow}) =
\left\{
\begin{array}{cl}
0 &  (\beta =  0) \\
\psi_{\uparrow\downarrow} & (\beta = \pi)
\end{array}
\right.
\end{equation}
and we then can write

\begin{equation} \label{5e22}
\begin{array}{rcl}
\psi^{(+)}_{\uparrow} & = & \sqrt{\rho} e^{-\gamma_{21} m t} \, ;  \\
\psi^{(+)}_{\downarrow} & = &  \sqrt{\rho} \gamma_{31} e^{-\gamma_{21}
m t} \, ; \\
\psi^{(-)}_{\uparrow} & = & \sqrt{\rho} \gamma_5 e^{-\gamma_{21} m t
}\, ; \\
\psi^{(-)}_{\downarrow} & = & \sqrt{\rho} \gamma_5 \gamma_{31}
e^{-\gamma_{21} m t} \, .
\end{array}
\end{equation}

The interesting point concerning these solutions is that here the
interpretation of $\proje_{\pm}$ as energy projector operators has no
meaning. Indeed, if we recall that the Tetrode (energy-momentum 1-form)
for the NLDHE is equal to the same tensor for the DHE \cite{vaz93},
i.e.

\begin{equation} \label{5e23}
T_\mu = \langle \dirac \psi \gamma_{210} \tilde{\psi} \gamma_{\mu}
\rangle \; ,
\end{equation}
we have  for all solutions (\ref{5e22}) that $T= T_{\mu}^{\mu} =
T_{\mu}.\gamma^{\mu} > 0$. Indeed,

\begin{equation}
T = \langle \dirac \psi \gamma_{210} \tilde{\psi}  \rangle_0 \; ,
\end{equation}
and using eq.(\ref{5e23}) we find

\begin{equation}
T = m \langle \psi \tilde{\psi} e^{-\gamma_5 \beta} \rangle_0 = m \rho
\, .
\end{equation}

On the other hand, in the case of the DHE the trace of the Tetrode
tensor is

\begin{equation}
T_{\mbox{Dirac}} = m \langle \psi \tilde{\psi} \rangle = m \rho \cos
\beta
\end{equation}
and we have

\begin{equation}
T_{\mbox{Dirac}} = T \cos \beta \, .
\end{equation}

We see that according to the NLDHE all solutions given by eq.\ref{5e22}
have positive energy and so in this theory it is a nonsense to
interpret $\proje_{\pm}$ as energy projector operators. In \cite{vaz93}
it is shown that $T_{\mu}$ can be interpreted as the ``extremal square
root'' of the Maxwell energy-momentum 1-form, $\displaystyle S_{\mu} =
\frac{1}{2} F \gamma_{\mu} F$,  which is a very interesting result.

Since we cannot interpret $\proje_{\pm}$ as energy projector operators
for the NLDHE we propose to interpret them as operators of particle
($\proje_+$) and antiparticle ($\proje_-$). In this way the
superposition principle continues to hold good only for solutions with
the same $\beta$. Explicitly this means that the superposition
principle holds for particles and for antiparticles separately, where
$\beta =0$ refers to particles and $\beta = \pi$ refers to
antiparticles. In this way the transformation $\beta \mapsto \beta +
\pi$ can be interpreted as transforming particle in antiparticle and
vice-versa.

To sum up, we saw that for consistence the DHE implies $\beta =  0$ or
$\beta = \pi$ for a free particle. On the other hand in  the NLDHE
$\beta=0$ and $\beta=\pi$ appear as conditions for $\psi$ to be an
eigenfunction of $\proje_{\pm} (\psi) = 1/2 (\psi + \gamma_0 \psi
\gamma_0)$. This suggests to us to search for  a new projection
operator which imply in other possible values for $\beta$ and in this
case these different values may eventually be associated with the other
known leptons. The new projection operator $\proje_\beta$ is introduced
by

\begin{equation}
\proje_\beta (\psi) = \frac{1}{2} (\psi + e^{\gamma_5 \beta} \gamma_0
\psi \gamma_0) \, .
\end{equation}
Then

\begin{equation}
\proje_+ = \proje_{\beta = 0} \; \mbox{and} \; \proje_- =
\proje_{\beta=\pi} \, .
\end{equation}
It is easy to verify that $\proje_\beta$ is indeed a projection
operator for all $\beta$ because

\begin{equation}
\proje_{\beta}^{2} = \proje_{\beta} \, , \; \proje_{\beta} =
\proje_{\beta + \pi} \, , \; \proje_{\beta} \proje_{\beta + \pi} =
\proje_{\beta + \pi} \proje_{\beta} = 0 \, .
\end{equation}

\subsection{Gauge Invariance of the Dirac-Hestenes Equation}

Consider the Dirac-Hestenes field in interaction with the
electromagnetic field $A \in \sec \bigwedge^1 (M) \subset \sec \clif
(M)$, which is described by

\begin{equation}
\dirac \psi \gamma_{21} + m \psi \gamma_0 + e A \psi =0 \; .
\end{equation}

It is well known that this equation is invariant  under the following
gauge transformations:

\begin{equation}
\psi \mapsto \psi \exp (e \gamma_{21} \theta) \; ; \quad A \mapsto A +
 \dirac \theta \; ;
\end{equation}
where $\theta \in \sec \bigwedge^0 (M) \subset \sec \clif (M)$.

Now let $\psi_+$ and $\psi_-$ be Weyl spinor fields, i.e. $\gamma_5
\psi_+ = + \psi_+ \gamma_{21}$, $\gamma_5 \psi_- = - \psi_-
\gamma_{21}$. Then $\psi_+$ and $\psi_-$ satisfy the Weyl equation

\begin{equation}
\dirac \psi_\pm = 0 \; .
\end{equation}

Consider the equation for $\psi_+$ coupled with an electromagnetic
potential $B$, i.e.

\begin{equation}
\dirac \psi_+ \gamma_{21} + g B \psi_+ = 0 \; .
\end{equation}
This equation is invariant under the gauge transformations

\begin{equation}
\psi_+ \mapsto \psi_+ \exp (g \gamma_{5} \theta) \; ; \quad B \mapsto B
+   \dirac \theta \; .
\end{equation}

On the other hand, the equation for $\psi_-$,

\begin{equation}
\dirac \psi_- \gamma_{21} + g B \psi_- = 0 \; ,
\end{equation}
is invariant under the gauge transformations

\begin{equation}
\psi_- \mapsto \psi_- \exp (-g \gamma_{5} \theta) \; ; \quad B \mapsto
B +   \dirac \theta \; .
\end{equation}

Consider now the Dirac-Hestenes spinor fields (eqs. \ref{2e17})

\begin{center}
\hfill
$\begin{array}{c}
\psi^\uparrow = \gamma_0 \psi_- \gamma_0 - \psi_- = \psi_{+}^{\uparrow}
- \psi_{-}^{\uparrow} \; ; \\
\psi^\downarrow = \psi_+ + \gamma_0 \psi_+ \gamma_0  =
\psi_{+}^{\downarrow} - \psi_{-}^{\downarrow} \; .
\end{array}
$
\hfill
(2.16')
\end{center}
which satisfy $P \psi^\uparrow = \psi^\uparrow$, $P \psi^\downarrow = -
\psi^\downarrow$. When $\psi^\uparrow$ is coupled with an
electromagnetic potential $B$ we have

\begin{equation}
\dirac \psi^\uparrow \gamma_{21} + g B \psi^\uparrow =0 \; ,
\end{equation}
which decouples in

\begin{equation}
\dirac \psi_{+}^{\uparrow} + g \gamma_5 B \psi_{+}^{\uparrow} = 0 \; ;
\quad \dirac \psi_{-}^{\uparrow} - g \gamma_5 B \psi_{-}^{\uparrow} = 0
\; .
\end{equation}

These results show clearly that $\psi^\uparrow$ describes a pair of
particles with charges $+g$ ($\psi_{+}^{\uparrow}$) and $-g$
($\psi_{-}^{\uparrow}$). We will call such particles magnetic monopoles
following the important work of \cite{locha85}, where similar ideas
have been introduced.

Now, writing $\psi_- = -S + {\cal F} - \gamma_5 P$ we have

\begin{equation}
\psi^\uparrow = \gamma_0 (-S + {\cal F} - \gamma_5 P) \gamma_0 - (-S +
{\cal  F} - \gamma_5 P) \; .
\end{equation}
Then from the fact that $\dirac \psi^\uparrow = 0$ we get

\begin{equation}
\dirac F^{\uparrow} = - \gamma_5 \dirac P \; ; \quad F^{\uparrow} =
\frac{1}{2} ( \gamma_0 {\cal F} \gamma_0 - {\cal F})\; ;
\end{equation}
which means that $\dirac \psi^\uparrow = 0$ is equivalent to a Maxwell
equation with magnetic current $J_m = \dirac P$, i.e.

\begin{equation} \label{5e43n}
\dirac F^{\uparrow} = - \gamma_5 J_m \; .
\end{equation}

\section{The Lepton Mass Spectrum}

We can now present our theory of the Lepton Mass Spectrum. We start by
supposing that the electron corresponds to a free electromagnetic field
configuration $F_e$, $\dirac F_e =0$ and $F_e^2 \neq 0$, such that for

\begin{equation}
F_e = \psi_e \gamma_{21} \psi_e
\end{equation}
we have
\begin{equation}
\dirac \psi_e \gamma_{21} + m \psi_e \gamma_{21}  = 0 \, ,
\end{equation}
which follows from the NLDHE once $\beta =0$. (For the positron we have
$\beta = \pi$.)

We now imagine that the other leptons are also electromagnetic
excitations. Since it is a known fact that, e.g. the muon decays
according to $\mu \rightarrow e + \nu + \overline{\nu}$ and that the
$\mu$ has the same intrinsic parity as $e^-$, we develop the following
idea.
We regard the muon as the electromagnetic configuration $F_e +
F^\uparrow = F$. From now on we
must use correct gaussian units as discussed above and
we must pay attention to the dimensions of the physical quantities. We
use
$[X]$ as meaning the physical dimension of a given quantity $X$.
$L$ means the length unit, $T$ the time unit, $M$ the mass unit and $Q$
the charge unit.
In gaussian units $[F] = QL^2$. Then, when we write $F=\psi \gamma_{21}
\tilde{\psi}$
we are taking $[\psi] = (QL^2)^{1/2}$. But usually $[\psi] =
L^{-3/2}$.
 Now, $[F] = QL^{-2} = Q (ML^2T^{-1})M^{-1}(LT^{-1})^{-1} L^{-3}$.
 Since $[e]=Q$,
$[\hbar] = (ML^2T^{-1})$, $[m] = M$ and $[c] = LT^{-1}$ we see that if
we take

\begin{equation} \label{4e9}
F = k \frac{e\hbar}{mc} \psi \gamma_{21} \tilde{\psi} \, ,
\end{equation}
where $k$ is a numerical constant and $m$ is the electron mass, the
units of $\psi$
result equal to $L^{-3/2}$. We show below that fixing $\displaystyle  k
= \frac{2\pi}{3}$
we can get the muon mass spectrum. We take $\psi^\uparrow$ with the
dimensions of an electromagnetic field.

If gaussian units are used we must write $\dirac F = (4\pi/c) {\cal
J}$. Then using eq.(\ref{5e43n}) we have

\begin{equation}
\dirac F = \frac{4\pi}{c} {\cal J} = - \frac{4\pi}{c} \gamma_5 J_m  \,
.
\end{equation}

The corresponding NLDHE for $\psi$ is obtained as in section 4. We get

\begin{equation} \label{5e40}
\dirac \psi \gamma_{21} = \frac{K_1c}{\hbar} \psi \gamma_0 e^{\beta
\gamma_5} + \gamma_5
\frac{K_2 c}{\hbar} \psi \gamma_0 e^{\beta \gamma_5} +
\frac{e^{\beta\gamma_5}}{\rho}
\left(  \frac{3m}{e\hbar} \right) {\cal J} \psi \, .
\end{equation}
Writing

\begin{equation}
\psi = e^{\gamma_5 \beta /2 } \phi \, ,
\end{equation}
where $\phi = \sqrt{\rho}R$ and supposing $\beta = \mbox{constant}$ we
have

\begin{equation} \label{5e42}
\dirac \phi \gamma_{21} = \frac{K_1 c}{\hbar} \phi \gamma_0 + \gamma_5
\frac{K_2 e}{\hbar} \phi \gamma_0 +
\frac{e^{\gamma_5 \beta}}{\rho} \left( \frac{3m}{e\hbar} \right) {\cal
J} \phi
\end{equation}
and differently from the case ${\cal J} = 0$ the factor $e^{\beta
\gamma_5}$ is not factored out. Now

\begin{equation} \label{5e43}
e^{\gamma_5 \beta} {\cal J} =\sin \beta J_m - \gamma_5  \cos \beta J_m
\, .
\end{equation}

For the free solution $\dirac F =0$ we arrived at $\dirac \psi
\gamma_{21} + (mc/\hbar) \psi \gamma_0 = 0 $, which in (\ref{5e42})
means to make $K_2 =0$ (as already explained) and to put $K_1 =m$, the
electron mass.

We look now for solutions of (\ref{5e42}) such that

\begin{equation} \label{5e44}
J_m = c q_m \phi \gamma_0 \tilde{\phi} \,
\end{equation}
where $q_m$  denotes the  ``magnetic'' charge. Using eqs. (\ref{5e43})
and (\ref{5e44}) in (\ref{5e40}) we get

\begin{equation}
\dirac \phi \gamma_{21} = \frac{Mc}{\hbar} \phi \gamma_0 + \gamma_5
\frac{Nc}{\hbar} \phi \gamma_0 \, ,
\end{equation}
which is analogous to eq.(\ref{eq.62}) but where now

\begin{equation}
M = m + \frac{3m}{e} q_m \sin \beta  \, ;
\end{equation}
\begin{equation}
N = - \frac{3m}{e} q_m \cos \beta  \, .
\end{equation}
This equation reduces to a DHE if $N=0$ and we have

\begin{equation}
\dirac \phi \gamma_{21} = \frac{Mc}{\hbar} \phi \gamma_0 \, .
\end{equation}

If $N=0$ we must have

\begin{equation} \label{5e59}
q_m \cos \beta = 0 \; \Rightarrow \; \beta = \frac{\pi}{2} \, , \;
\beta = \frac{3\pi}{2} \, .
\end{equation}
(Before proceeding we suppose that to $\beta = \pi/2$ is associated
$+q_m$ and to $\beta = 3\pi/2$ is associated $-q_m$.)

With eq.(\ref{5e59}) we get

\begin{equation}
M =m + 3m \frac{q_m}{e} \, .
\end{equation}

It can be shown \cite{locha85} that the monopole theory developed in
section 5.2 implies in Dirac's quantization condition :

\begin{equation}
\frac{eq_m^n}{\hbar c} = \frac{n}{2}, \quad \mbox{$n$ integer} \, .
\end{equation}
where $e$ is the electric charge of the electron. Then

\begin{equation}
q_m^n = \frac{e}{2\alpha} n
\end{equation}
where $\alpha = e^2/\hbar c$ is the fine structure constant. Taking
$n=1$ we have

\begin{equation}
M = \left( 1 + \frac{3}{2\alpha} \right) m \, ,
\end{equation}
which gives $M = 206.55 m = 105.5 \,\mbox{MeV}$. The muon mass is
$m_\mu = 206.70 m$. We see that our theory gives an excellent result.

For $\beta = \pi/2$ or $\beta = 3\pi/2$ we have the following
solutions:

\begin{equation} \label{5e64}
\begin{array}{rcl}
\psi^{(+)}_{\uparrow} & = & \sqrt{\rho} e^{\gamma_5 \pi/4}
e^{-\gamma_{21} Mc t/\hbar } \, ;  \\
\psi^{(+)}_{\downarrow} & = &  \sqrt{\rho} e^{\gamma_5 \pi/4}
\gamma_{31} e^{-\gamma_{21} Mc t/\hbar} \, ; \\
\psi^{(-)}_{\uparrow} & = & \sqrt{\rho} e^{\gamma_5 3\pi/4}
e^{-\gamma_{21} Mct/\hbar } ) \, ; \\
\psi^{(-)}_{\downarrow} & = & \sqrt{\rho}  e^{\gamma_5 3\pi/4}
(\gamma_{12} e^{-\gamma_{21} Mct/\hbar} )\, .
\end{array}
\end{equation}
where the solutions with index $(+)$ are eigenspinors of $\proje_{\beta
= \pi/2}$ and the solutions with index $(-)$ are eigenspinors of
$\proje_{\beta = 3\pi/2}$, i.e.

\begin{equation}
\begin{array}{rcl}
\proje_{\beta = \pi/2} ( \psi^{(+)}_{\uparrow \downarrow} ) &=&
\psi^{(+)}_{\uparrow \downarrow} \, ; \\ & & \\
\proje_{\beta = 3\pi/2}( \psi^{(-)}_{\uparrow \downarrow} ) &=&
\psi^{(-)}_{\uparrow \downarrow} \, .
\end{array}
\end{equation}
In this way we may take the projector operators $\proje_\beta$ with the
appropriate values of $\beta$ as the projectors of electron, positron,
muon and antimuon, i.e.

\begin{equation}
\begin{array}{ll}
\proje_{\beta =0} = \proje_{e^-} \, ; & \proje_{\beta =\pi} =
\proje_{e^+} \, ; \\
\proje_{\beta =\pi/2} = \proje_{\mu^-} \, ; & \proje_{\beta =3\pi/2} =
\proje_{\mu^+} \, .
\end{array}
\end{equation}

At this point the question that naturally arises is the following: May
the above theory give the mass of other leptons, in particular the mass
of the {\em Tau}~? This can bi done using an {\em ad hoc\/}
hypothesis.
We suppose that the Tau is made of an excited state of the muon.
Proceeding as before we arrive at a new equation for $M$, which we now
write $\overline{M}$:

\begin{equation}
\overline{M} = \left( 1 + \frac{3}{2\alpha} \right) m + 3 m
\frac{q_m^n}{e} \, .
\end{equation}
In this equation, supposing $n = 2^4 = 16$ in the Dirac quantization
condition, we have

\begin{equation}
\overline{M} = \left( 1 + \frac{3}{2\alpha} \right) m +
\frac{3}{2\alpha} 2^4 m  = \left( 1 + 17 \frac{3}{2\alpha} \right) m =
3845 m = 1785 \; \mbox{MeV/$c^2$} \, ,
\end{equation}
which is a good approximation for the  Tau mass. The hypothesis $n=p^4$
($p$ integer) in the formula for the possible values of $q_m^n$ leads
to the spectrum

\begin{equation} \label{5e69}
M_p = m + \frac{3}{2\alpha} m \sum_{l=0}^{p} l^4 \, ,
\end{equation}
which corresponds to a formula found by \cite{barut80} based on
arguments completely different from the ones presented above. From
eq.(\ref{5e69}) we obtain $M_0 = m$, $M_1 = m_\mu$, $M_2=m_\tau$ and
$M_3 = 20090 m = 10291 \, \mbox{MeV/$c^2$}$ and such a lepton has not
been found as yet.

\section{Conclusions}

We saw in this paper that there are many news from Maxwell and Dirac.
The existence of subluminal and superluminal UPW solutions of the free
Maxwell equations looks, at least to the authors, an extraordinary fact
with implications in all branches of physics. For a discussion of these
implications see \cite{upw}. We saw that these solutions correspond to
solutions with field invariants different from zero and that they have
also longitudinal components. In this respect we must say that
\cite{eva94} presents some evidences that there are electromagnetic
waves with longitudinal electromagnetic fields. Also important is the
fact that Maxwell equations are equivalent to a NLDHE which, for
particular values of the Takabayasi angle, gives the lepton mas
spectrum through the construction of section 5.

We finish by commenting that Weyl equations also have subluminal and
superluminal solutions \cite{vazrod95,upw}. This may eventually explain
some of the mysteries associated to neutrinos, because in both cases
they move as massive particles with momenta satisfying $p_{<}^{2} =
\Omega^2 > 0$ and $p_{>}^{2} = -\Omega^2 < 0$, the symbols $<$ and $>$
corresponding to the subluminal and superluminal solutions.

\acknowledgements{The authors are grateful to Dr. Q.A.G. de Souza and
Mr. J.E. Maiorino for discussions, and to CNPq and FAPESP for partial
financial support.}

\end{document}